\documentclass[preprint]{aastex}

\newcommand{\BV}{Brunt-V\"ais\"al\"a \ }

\newcommand{\uu}{\mbox{\boldmath $u$} {}}
\newcommand{\kk}{\mbox{\boldmath $k$} {}}
\newcommand{\nab}{\mbox{\boldmath $\nabla$} {}}

\shorttitle{Convective Overstability in Accretion Disks}
\shortauthors{Klahr \& Hubbard}

\begin{document}


\title{Convective Overstability in radially stratified accretion disks under thermal relaxation}


\author{Hubert Klahr\altaffilmark{1} \& Alexander Hubbard\altaffilmark{2}}
\email{klahr@mpia.de}
\email{VERSION: Apr.1st 2014}
\email{ApJ in press}

\altaffiltext{1}{Max-Planck-Institut f\"ur Astronomie, K\"onigstuhl
                 17, 69117, Heidelberg, Germany;
                 klahr@mpia.de}
\altaffiltext{2}{Department of Astrophysics, American Museum of Natural History, New York, NY 10024-5192, USA }




\begin{abstract}
This letter expands the stability criterion for radially stratified, vertically {unstratified} accretion disks incorporating thermal relaxation. We find a linear amplification of epicyclic oscillations in these disks that depends on the effective cooling time,
 i.e.\ an overstability.  
The growth rates of the overstability vanish for both extreme cases, e.g.\ infinite cooling time and instantaneous cooling, i.e.\ 
the adiabatic and fully isothermal cases.
However, for thermal relaxation times $\tau$ on the order of the orbital frequency, $\tau\Omega \sim 1$, modes
grow at a rate proportional to the square of the \BV frequency.
The overstability is based on epicyclic motions, with the thermal relaxation causing gas
to heat while radially displaced inwards, and cool while radially displaced outwards.  This causes the gas to have a lower density
when moving outwards compared to when it moves inwards, so it feels the outwards directed pressure force more
strongly on that leg of the journey.
We suggest the term ``Convective Overstability" for the phenomenon that has already been numerically studied
in the non-linear regime in the context of amplifying vortices in disks,
under the name ``Subcritical Baroclinic Instability".
The point of the present paper is to make clear that vortex formation in {three-dimensional} disks is neither subcritical,
i.e.\ does not need a finite perturbation, nor is it baroclinic in the sense of geophysical fluid dynamics,
which requires on vertical shear. We find that Convective Overstability is a linear instability that will operate 
under a wide range of physical conditions for circumstellar disks.
\end{abstract}


\noindent


\section{Introduction}
The hydrodynamical stability of circumstellar accretion disks has been a long-standing problem in astrophysics because
turbulence appears to be needed to drive observed accretion flows \citep{SS73}.
However, these disks are strongly stabilized by rotation
and vertically stable stratification. While the identification of the role of Magneto-Rotational Instability in accretion disks (MRI, \citealt{Vel59,BH91}) 
finally provided an incontrovertible linear instability, it was soon noted \citep{Gammie96} that significant portions of
circumstellar disks are too poorly ionized to allow the MRI to act.
The  upper and
outer edges for MRI turbulence appear to be set by ambipolar diffusion
\citep{2011ApJ...727....2P, 2011ApJ...739...50B} and by the stiffness
of the fields at high magnetic pressure \citep{2000ApJ...534..398M, 2000ApJ...540..372K, 2010ApJ...708..188T}.
These obstacles of the otherwise robust MRI continue to motivate studies of hydrodynamical instabilities.

The radial temperature structure of modestly accreting circumstellar disks are largely controlled by irradiation from the central star, although
any accretion flow will lead to increased heating and dependent on the local opacity to a non trivial radial temperature and density profile \citep{bell95, D'Alessio05}.
The radial temperature gradient can overpower the expected density gradient, leading to a negative radial entropy gradient
\citep{KRL13}, which can drive the formation of vortices through a baroclinic mechanism, e.g. the non-vanishing baroclinic term in the vorticity equation \citep{KB03}.
Disks without thermal relaxation process appear to be linearly stable \citep{K04}, but
\citet{Petersen07} showed that vortices will form and grow if, on top of the radial entropy gradient,
there is also sufficiently fast thermal relaxation and a strong enough initial perturbation.
\citet{LP2010} and \citet{LK2011} presented 3D results of vortex
amplification for vertically unstratified disks with imaginary radial \BV (buoyancy) frequencies and short relaxation times on the order of  the orbital period.
\citet{LP2010} referred to this process as a Subcritical Baroclinic Instability (SBI) because it appeared that one needs finite size
perturbations to create the first vortices, which can then be amplified by a convective radial entropy flux.
Thus, while the problem of amplifying vortices to sizes and strengths where they could drive significant accretion flows even in
MRI-inactive regions was solved,
the question of the origin of vortices remained open.

In the present paper we perform a linear stability analysis for radially stratified accretion disks. 
{The results are not suited to explain the results for 2D vertically integrated accretion disks as presented in \citet{Petersen07} and \citet{RLK13} because we need the vertical dimension to achieve pressure equilibrium.  However, this paper explains the behavior of
the 3D yet vertically unstratified disk models \citep{LP2010,LK2011}.}

Vertical stratification, {on the other hand,}
is accompanied by vertical shear, which introduces additional potential sources of instability in a flow, e.g. the Goldreich-Schubert-Fricke instability.
 A full 3D stratified analysis is under way and shall be presented in a separate paper.

We will start in section 2 with the linear analysis for an anelastic ansatz, e.g. a disk without pressure fluctuations in which density is a pure 
function of the background pressure and the local temperature.  {In this approximation,} the continuity equation {is incorporated in the equation for the specific entropy.} 
In section 3 we compare with numerical and previous results.  Our code and numerical setup is described in the Appendix.
We conclude in section 4.

\section{Linear Stability Analysis}

\subsection{Quasi hydrostatic approximation}

We consider the inviscid hydrodynamic equations, including thermal relaxation, in polar coordinates $(R, \phi)$ for a vertically {unstratified } disk. 
Compressibility effects such as sound waves do not play a role in
vortex formation, so we adopt a {special} form of the anelastic approximation: the pressure $p(R)$ does not change.
{This requires vertically thin perturbations which can equilibrate pressure vertically on sound crossing times, far faster than
the system's dynamical time scale.}
This means that the density can be calculated from the background pressure and the local
 specific entropy using the relation ${\rm d} S = C_{\rm v} {\rm d} \log (p \rho^{-\gamma})$. The adiabatic index $\gamma$ for circumstellar material (e.g. a mixture of Hydrogen and Helium) in 3D is $1.43$.

{{Our approximation} enables us to neglect the continuity equation, but requires that the 3D divergence of velocity stays small\footnote{More precisely in $\nab \cdot  \rho \uu$ has to remain small.}.
More precisely the divergence of the flow field $\nab \cdot \uu$ has to be small in order that the Eulerian pressure perturbations can be neglected:
\begin{equation}
\nab \cdot \uu \approx -\gamma^{-1} \uu \nab \left(\ln p_0\right) \ll |\kk|  |\uu|. \label{div_u_eq}
\end{equation}
with $\kk$ being the WKB wave-vector. 
This implies to the leading order in the ratio of the WKB wavelength to the radial pressure scale length
\begin{equation}
k_R u_R + \frac{m}{R} u_\phi + k_z u_z \approx 0,
\end{equation}
where $m$ is the azimuthal wavenumber.
There are now two distinct scenarios that {could} fulfill this criterion
($k_z=0$ and $u_z=0$ reduce to the same system).
If $u_z = 0$ then Eq.~\ref{div_u_eq} reduces to $u_r \approx -\frac{m}{R} u_\phi \ll u_\phi$, which means that
$k_r R/m \gg 1$ and the modes must be strongly elongated in the azimuthal direction.  This condition will be invoked
 later in the tight winding assumption to be able to treat $k_r$ and $\omega$ as slowly varying in time.
Significantly, the initial vortices one finds in non-linear simulations of radial stratified accretion disks (e.g. Raettig et al.\ 2013), which tend to grow slowly by the mechanism described in in Lesur \& Paploizou (2010), are in fact extremely elongated in the azimuthal direction.
The large $u_\phi$ of an elongated vortex means that perturbing the azimuthal velocity is energetically expensive compared to perturbing
the radial velocity.  It follows that strongly elongated modes, which require large energy inputs, will grow slower than derived in this paper for epicyclic motion {and are not a suited example for the epicyclic motions discussed in this paper.}

{We focus on} the other possibility is that $v_z \ne 0$, allowing {for the readjustment of the hydrostatic equilibrium} to be maintained by vertical motions.
Then we have $k_z \gg k_R, m/R$ as our incompressibility condition with $u_R \gg u_z$.  This implies vertical layers of thin sheets of gas that oscillate against each other, a fact that we were able to confirm in our numerical simulations (see Fig.\ 4). This situation is very similar to the ideal nonaxisymmetric MRI  modes in the presence of a mean toroidal field (Balbus and Hawley 1992). There too, the most strongly amplified perturbations are those for which $k_z \gg k_R, m/R$ during the period of transient exponential growth, because horizontal compressions are not resisted by horizontal pressure gradients.


\subsection{Instability Analysis}

The equations for radial, vertical and azimuthal velocities $u_R, u_z, u_\phi$ are:
\begin{equation}
\partial_t u_R + u_R \partial_R u_R + \frac{u_\phi}{R}\partial_\phi u_R + u_z \partial_z u_R - \frac{u_\phi^2}{R} = - \frac{1}{\rho}\partial_R p + g_R
\label{uR1}
\end{equation}
\begin{equation}
\partial_t u_z + u_R \partial_R u_z + \frac{u_\phi}{R} \partial_\phi u_z  + u_z \partial_z u_z = - \frac{1}{\rho}\partial_z p
\label{uz1}
\end{equation}
\begin{equation}
\partial_t u_\phi + u_R \partial_R u_\phi + \frac{u_\phi}{R}\partial_\phi u_\phi + u_z \partial_z u_\phi + \frac{u_\phi u_R}{R} = - \frac{1}{R \rho}\partial_\phi p,
\label{uphi1}
\end{equation}
where $g_R$ is the radial component of stellar gravity.
We use an entropy formulation for the energy equation:
\begin{equation}
\partial_t S + u_R\partial_R S + \frac{u_\phi}{R} \partial_\phi S + u_z\partial_z S = -\frac{C_{\rm v}}{T}\frac{T - T_0}{\tau}.
\end{equation}
The relaxation term on the right side reestablishes the original temperature (entropy) profile of the disk after entropy is radially transported. We can replace temperature in this equation by entropy for small perturbations in temperature $T_1 = T - T_0$ and pressure $p_1 = p - p_0$:
\begin{equation}
\partial_t S = \partial_t \left( C_{\rm v} \left[(1-\gamma)\frac{p_1}{p_0} + \gamma\frac{T_1}{T_0} \right] \right),
\end{equation}
and because the pressure is fixed here it follows $p_1 = 0$:
\begin{equation}
\partial_t S = \partial_t \left(C_{\rm v} \left[\gamma\frac{T_1}{T_0} \right]\right),
\end{equation}
Effectively the thermal relaxation time $\tau$ can be translated
into an entropy relaxation time $\tau_{\rm S} = \gamma\tau$. 
Thus the entropy equation simplifies to:
\begin{equation}
\partial_t S + u_R\partial_R S + \frac{u_\phi}{R} \partial_\phi S + u_z \partial_z S = -\frac{S - S_0}{\gamma\tau}.
\label{S1}
\end{equation}

Next we linearize using the ansatz $u_R = u_R', u_z = u_z', u_{\phi} = u'_{\phi} + \Omega R, S = S_0 + S_1$, and we drop the primes for the velocity
perturbations.
Note that $p'$ is zero under {the quasi hydrostatic} condition,
and that background quantities are axisymmetric.  Eqs \ref{uR1}, \ref{uz1}, \ref{uphi1} and \ref{S1} become
\begin{equation}
\partial_t u_R + \Omega \partial_\phi u_R - 2\Omega u_\phi - \frac{\rho_1}{\rho_0^2}\partial_R p_0 = 0,
\label{uR2}
\end{equation}
\begin{equation}
\partial_t u_\phi +\frac{u_R}{R}\partial_R(R^2 \Omega) + \Omega \partial_\phi u_\phi  = 0,
\label{uphi2}
\end{equation}
\begin{equation}
\partial_t u_z + \Omega \partial_\phi u_z  = 0
\label{uz2}
\end{equation}
\begin{equation}
\partial_t S_1 + u_R\partial_R S_0 + \Omega \partial_\phi S_1
  = - \frac{S_1}{\gamma\tau}.
\label{S2}
\end{equation}
As we already noted, the density $\rho$ depends only on entropy for locally constant pressure, so
the entropy fluctuation $S_1$ can expressed as:
\begin{equation}
S_1 = C_{\rm v} \left[-\gamma\frac{\rho_1}{\rho_0} \right].
\end{equation}
Thus we can recast the entropy equation as an evolution equation for the density :
\begin{equation}
\gamma \left(-\partial_t - \Omega \partial_\phi - \frac{1}{\gamma\tau}\right) \frac{\rho_1}{\rho_0} + u_1\frac{1}{C_{\rm v}}\partial_R S_0 = 0.
\end{equation}

Following \cite{Rudiger02}, we perform a local analysis in a small volume around the point $(R_0, \phi_0)$, and thus the coefficients in the above equation are constant. We also apply the short wave and small $m$ approximation, e.g.\ $|k_z z|>>|k R|>>m$, which is effectively a WKB approach. We write our perturbation as $\exp[i(kR + m\phi + k_z z- \omega t)]$ and arrive at the following system of equations:
\begin{equation}
-i (\omega-m\Omega) u_R  - 2\Omega u_\phi- \frac{\rho_1}{\rho_0^2}\partial_R p_0 = 0,
\end{equation}
\begin{equation}
-i (\omega-m\Omega) u_\phi +\frac{u_R}{R}\partial_R(R^2 \Omega) = 0,
\end{equation}
\begin{equation}
-i (\omega-m\Omega) u_z = 0
\end{equation}
and
\begin{equation}
\left(i \omega - i m \Omega - \frac{1}{\gamma\tau}\right) \frac{\rho_1}{\rho_0} + u_1 \frac{1}{\gamma C_{\rm v}}\partial_R S_0  = 0.
\end{equation}
{It is interesting to note that the vertical velocity is independent from the other three equations and
therefore does not have to be considered any further for the purpose of investigating stability.
In the final solution the evolution of $u_z$ and corresponding wave number $k_z$ would be determined via the $p_1= 0$ condition, see discussion above.}
We replace $\omega - m \Omega$ by $\omega_m$ and write our matrix as
\begin{mathletters}
\begin{equation}
\left|
\begin{array}{ccc}
-i\omega_m  & -2 \Omega &-\frac{1}{\rho_0}\partial_R p_0\\
 \frac{1}{R}\partial_R(R^2 \Omega)&-i \omega_m & 0\\
\frac{1}{\gamma C_{\rm v}}\partial_R S_0 & 0 &+i\omega_m -\frac{1}{\gamma\tau}\\
\\
\end{array} \; \; \right|\;
\;  \;
\end{equation}
\end{mathletters}
For non-trivial solutions, the determinant of the matrix has to vanish, which leads to the dispersion relation:
\begin{equation}
(\omega_m^2-\kappa_R^2)\left(\omega_m + \frac{i}{\gamma\tau}\right) - \omega_m N_R^2= 0,
\label{disp_rel}\end{equation}
in which we use as abbreviations the epicyclic frequency $\kappa_R^2 = \frac{1}{R^3}\frac{\partial \Omega^2 R^4}{\partial R}$, which 
measures the gradient in specific angular momentum; and the radial buoyancy frequency $N_R^2 = -\frac{1}{\rho_0 \gamma C_{\rm v}}\partial_R p_0\partial_R S_0$.


In the fast cooling limit, $\tau = 0$, we recover the Rayleigh criterion in rotating flows:
\begin{equation}
\omega_m^2 = \kappa_R^2.
\end{equation}
In the adiabatic limit, $\tau = \infty$, we find the Solberg-Hoiland Criterion for unstratified disks:
\begin{equation}
\omega_m^2 = \kappa_R^2 + N_R^2.
\end{equation}
However, systems that are stable according to the Solberg-Hoiland Criterion can still be unstable (or rather, overstable)
 in the presence of finite thermal relaxation
($0 < \gamma\tau \Omega < \infty$).
To calculate the growthrates we write $\omega_m = \omega + i \Gamma$, with $\omega$ and $\Gamma$ real. 
Then $\omega_m^2 = \omega^2 - \Gamma^2 + 2 i \omega \Gamma$ and thus 
\begin{equation}
(\omega^2 - \Gamma^2 + 2 i \omega \Gamma-\kappa_R^2)\left(\omega + i \Gamma + \frac{i}{\gamma\tau}\right) - ( \omega + i \Gamma) N_R^2= 0.
\end{equation}
Sorting for real and imaginary parts, which have to vanish independently, we can determine $\Gamma$:
\begin{eqnarray}
\Gamma\left(2\Gamma + \frac{1}{\gamma\tau}\right)^2 + \Gamma(\kappa_R^2 + N_R^2) + \frac{1}{2\gamma\tau} N_R^2= 0, 
&\rm{for~}\omega \neq 0, \label{omega_finite}\\
\left(\Gamma^2 +\kappa_R^2\right)\left(\Gamma + \frac{1}{\gamma \tau}\right)+\Gamma N_R^2 =0,  &\rm{for~}\omega=0.
\label{omega_zero}
\end{eqnarray}
Note that $\omega \neq 0$ implies a stable or damped oscillation ($\Gamma \leq 0$) or a growing oscillation, i.e.~an
overstability ($\Gamma>0$). {In the case of $\tau \omega >> 1$ the oscillation frequency is approximately given by $\omega \approx \pm \sqrt{\kappa_R^2 + N_R^2}$, i.e.~the epicyclic frequency.}
If  $\omega=0$ the system is either conventionally stable ($\Gamma \leq 0$) or unstable ($\Gamma>0$).

Eq.\ \ref{omega_finite} has two complex roots with negative real parts, which are not relevant for the problem here.
The third root is real and can easily be approximated if one assumes that the growth rate $\Gamma$ is smaller than the thermal relaxation rate
$\tau^{-1}$: 
\begin{equation}
\Gamma = \frac{1}{2}\frac{- \gamma\tau N_R^2}{1+\gamma^2 \tau^2\left(\kappa_R^2 + N_R^2\right) },
\label{GAMMAEQ}
\end{equation}
which demonstrates linear overstability if $N_R^2<0$.
This approximation holds even for buoyancy frequencies close to the Keplerian frequency, which is 
not a relevant physical condition in circumstellar disks because in general $|\kappa_R| \gg |N_R|$.

In the other extreme of $\omega=0$ (i.e.~an instability, instead of an overstability) and
growth rates much larger than
the relaxation rate, i.e. the adiabatic limit, we again recover the Solberg-Hoiland result:
\begin{equation}
\Gamma^2 = - \left(\kappa_R^2 + N_R^2\right),
\end{equation}
the classical growth rate for convection in adiabatic rotating systems.

\begin{figure}
\includegraphics[width=\columnwidth]{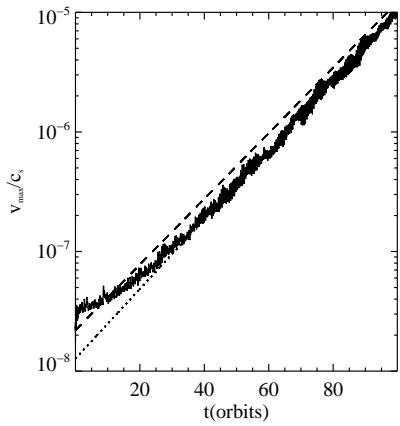}
\caption{Analytical predicted growth of perturbations for convective overstability and numerically determined values.
The dotted line is the fit of the last 50\% of the data, and the dashed line is the growth as analytically determined for the inner edge
of the measurement volume inside the simulation region ($R = 0.9$).
The initial perturbation was $\pm 10^{-8}$ times the speed of sound. The radial powerlaws of temperature
and density are q = -2.0 and p = -1.32. 
\label{Fig_thov0.eps}}
\end{figure}

\begin{figure}
\includegraphics[width=\columnwidth]{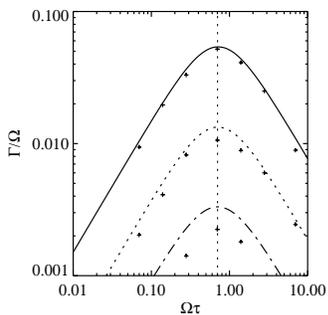}
\caption{Analytical predicted growth-rates for convective overstability and numerically determined values.
Solid line is for q = -2.0 and p = -1.32, dotted line is for q = -1.0 and p = -0.66 and dash dotted line is for q = -0.5 and p = -0.33. Crosses indicate numerical determined growth rates from direct numerical simulations. The vertical line indicates the theoretical maximum of the curves at $\gamma \Omega \tau = 1$.
\label{Fig_thov1.eps}}
\end{figure}

\begin{figure}
\includegraphics[width=\columnwidth]{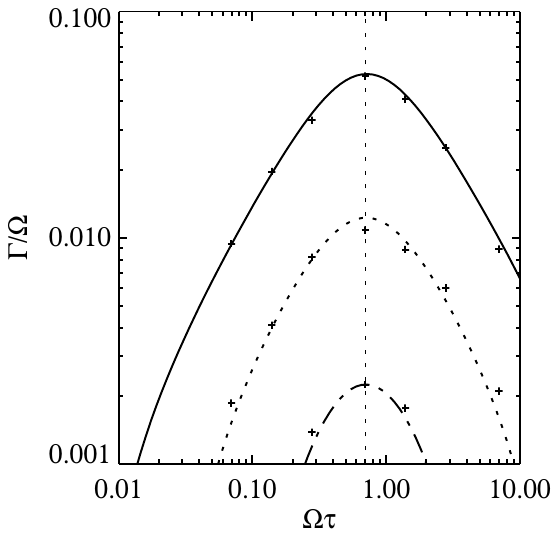}
\caption{As Fig.\ \ref{Fig_thov1.eps}, except that the analytical growth rates have been reduced by $0.001 \Omega$, approximating the
numerical dissipation.
\label{Fig_thov2.eps}}
\end{figure}

\section{Comparison to numerical and previous results}

\subsection{Numerical results}

In Fig.\ \ref{Fig_thov0.eps} we plot the maximum velocity in a very weakly initialized (perturbation amplitude of $10^{-8} c_s$) numerical simulation with
density and temperature powerlaws $p \equiv \frac{d {\rm \log} \rho}{d {\rm \log} R}=-1.32$ and $q \equiv \frac{d {\rm \log} T}{d {\rm \log} R} =-2$
(details in the appendix).
The dotted line is the fit to the last 50 orbits of data, while the dashed line shows the growth rate as analytically predicted
at the inner edge of the measurement volume.  We can see that the analytical estimate fits the data quite well.

In Fig.\ \ref{Fig_thov1.eps} we plot the growth rate (Eq.\ \ref{GAMMAEQ}) for a range of circumstellar disk parameters, i.e. we vary the radial density 
power law $p= \left[-0.33, -0.66, -1.32 \right]$ and $q =\left[-0.5, -1, -2 \right]$.  
Please note that typical radial temperature gradients in observed disks around young stars should fall in the range  of $q = -0.6 \pm  0.2$ \citep{andrews10}, where as the radial density gradient (as opposed to surface density gradient) is a function of height in the disk above midplane, and thus can easily also obtain the local values as proposed above.

The perturbation is stronger than
in Fig.\ \ref{Fig_thov0.eps}, with an amplitude of $10^{-4} c_s$ for faster convergence of the models, especially for the weakly stratified cases. 
The pressure scale height chosen here is $H/R = 0.1$, and the epicyclic frequency is very nearly $\kappa_R \simeq \Omega \gg N_R$.
Note that $N_R^2<0$ requires $(1-\gamma)p+q<0$, which is fulfilled for all three of our cases.
The growth rates reach their maximum for $\gamma\tau \Omega = 1$, and the maximum growth rates are 
\begin{equation}
\Gamma = -\frac{1}{4} N^2/\Omega = -\frac{3}{8} Ri \Omega. \label{E24}
\end{equation}

While the analytical estimate overstates the growth rate in the latter two curves, their growth rates are low enough
that numerical viscosity plays a role.   In that case, Eq. \ref{GAMMAEQ} becomes
\begin{equation}
\Gamma = \frac{1}{2}\frac{- \gamma\tau N_R^2}{1+\gamma^2 \tau^2\left(\kappa_R^2 + N_R^2\right) } - \frac{\nu}{l^2},
\label{GAMMAEQDIF}
\end{equation}
where $\nu$ is a dissipation coefficient that includes numerical viscosity.
In Fig.\ \ref{Fig_thov2.eps} we show the same data as in Fig\ \ref{Fig_thov1.eps},
except that we plot Eq. \ref{GAMMAEQDIF} with $\nu/l^2=0.001 \Omega$, demonstrating an excellent fit.

\subsection{Comparison to previous results}

The cooling time can be approximated from the underlaying radiation energy diffusion process.
Thermal relaxation times over one disk pressure scale height via radiative processes are of the order
$1 < \gamma\tau(H) \Omega < 100$ \citep{KRL13} and are calculated using the effective temperature diffusion parameter $\mu = \frac{\lambda 4 c a_R T^3}{\rho \kappa}$ (see for instance \citet{bitsch09}) for typical dust opacities in circumstellar disks (Bell \& Lin 1997).
We can then derive the cooling time over lengthscale $l$ as 
\begin{equation}
\tau(l) = \frac{l^2}{\mu}.
\label{tau}
\end{equation}
The condition $\tau \gamma \Omega = 1$ can be interpreted as Peclet number of $1/\gamma$ because
$Pe = l^2\Omega/\mu = \tau \Omega$.

Plugging Eq.\ \ref{tau} into Eq.\ \ref{GAMMAEQDIF},
we arrive a very similar expression to Eq.\ (22) in \citet{LP2010}, which they derived as estimates for vortex growth. 
Our equation reads as 
\begin{equation}
\Gamma = \frac{1}{2}\frac{- \frac{l^2}{\mu} \gamma N_R^2}{1+\left(\gamma\frac{l^2}{\mu}\right)^2\left(\kappa_R^2 + N_R^2\right)} -\frac{\nu}{l^2},
\label{GAMMALP0}
\end{equation}
whereas \cite{LP2010} derived
\begin{equation}
\Gamma = - \frac{l^2}{\mu} N_R^2 \Phi_\omega(S \frac{l^2}{\mu}) -\frac{\nu}{l^2},
\label{GAMMALP1}
\end{equation}
leaving the phasing mixing term $\Phi_\omega$ undefined. Both results are identical if 
one set the phase mixing term to
\begin{equation}
\Phi_\omega(S \frac{l^2}{\mu}) = \frac{1}{2}\frac{\gamma}{1+\left(\gamma\frac{l^2}{\mu}\right)^2\left(\kappa_R^2 + N_R^2\right) }.
\label{GAMMALP2}
\end{equation}
As already pointed out by \cite{LP2010}, one finds the Rayleigh criterion for $\tau \Omega = \frac{l^2}{\mu} \Omega = Pe << 1$
because the growth rates are:
\begin{equation}
\Gamma = -\frac{1}{2} \frac{l^2}{\mu}\gamma  N_R^2 -\frac{\nu}{l^2},
\label{GAMMALP3}
\end{equation}
which leads to the condition for growth
\begin{equation}
- \frac{l^4}{\nu\mu} N_R^2 = Ra > \frac{2}{\gamma}.
\label{GAMMALP4}
\end{equation}
Interestingly the critical Rayleigh number for convective overstability is only $2/\gamma$ instead of $Ra_c =1708$ for Rayleigh Bernard Convection. 
This is now the condition that the smallest epicyclic perturbations or vortices get amplified, because they are the strongest affected by thermal relaxation. We can rescale Eq.\ \ref{GAMMALP4} to the values at the pressure scale height $H$ using
\begin{equation}
Ra = - \frac{l^4}{H^4}\frac{H^4\Omega^2}{\nu\mu} Ri =  - \frac{l^4}{H^4} Pe Re Ri
\label{GAMMALP5}
\end{equation}

\subsection{New results}
With the complete relation of the growth behavior we can also derive
general criteria for growth if the the cooling time is not extremely short.
From Eq. (\ref{E24}) it follows that growth in the perfect regime $\gamma \tau \Omega = 1$ only occurs if
\begin{equation}
-\frac{3}{8} Ri \Omega > \frac{\nu}{l^2}\label{E25}
\end{equation}
which can be translated into a Reynolds number Criterion:
\begin{equation}
Re > \frac{8}{3 Ri} \label{E26}
\end{equation}
under the additional condition for the Prandtl number that states
the ratio between the viscosity and the heat transfer rate of a fluid, which is given by the ratio of
Peclet number over Reynolds number. Thus the critical Prandtl number is
\begin{equation}
Pr = Pe/Re = \frac{3}{8 \gamma} Ri.
\end{equation}
It follows that to study the the Convective Overstability experimentally, one will have to 
focus on fluids that have a Pr well below unity; because a fluid with a larger 
Prandtl number needs a Richardson number larger that 1, in which case one
is no longer in the overstable but in the linearly convective unstable regime
as indicated by the Solberg-Hoiland criterion.

For cooling times longer than the orbital period the conditon is rendered by
\begin{equation}
\frac{1}{2} \frac{-N_R^2}{\kappa_R^2 + N_R^2} \gamma\frac{\mu} {l^2} > \frac{\nu}{l^2},
\label{GAMMALPX1}
\end{equation}
and thus the needed minimum Prandtl number for instability is identical to the one obtained in the optimal case explained above.

\section{Conclusions}
We were able to put forward a linear theory for instability in accretion disks which are radially stratified and subject to radiatively driven thermal relaxation. This is the first analysis to our knowledge that incorporates finite thermal relaxation into the Solberg Hoiland Criteria for rotating fluids as are accretion disks. For realistic parameter ranges of the radial temperature gradient (see \citet{andrews10}) and thermal relaxation \citep{KRL13}, we find the amplification of radial epicyclic oscillations on a time scale of 100 to 1000 orbits. 

We tested our analytic approximations  by comparing the results to numerical simulation of the growth of small perturbations in cylindrical unstratified and axis symmetric accretion disks. {Most importantly we tested our assumption about the fixed pressure background a posteriori. Pressure fluctuation are measured to be an order of magnitude lower than adiabatic pressure variations following from local compression, e.g. density fluctuations.} The compliance of numerical and analytical results is striking, especially when one takes the numerical viscosity of the code into account.  Saturation values for the axissymetric and non-axissymmetric cases shall be obtained in future simulations at higher resolution studies,
which hopefully will be less hampered by numerical dissipation.

As a result we have shown that the Subcritical Baroclinic Instability is {in 3D simulations} not necessarily subcritical, nor a baroclinic instability in the traditional sense, nor any kind of  stationary instability, but an overstability. 
As an excerpt from Chandrasekhar's book (1961) we quote: `Eddington explains this choice of terminology as follows: ``In the usual kinds of instability, a slightly displacement provokes restoring forces tending away from equilibrium; in an overstability it provokes restoring forces so strong, as to overshoot the corresponding position on the other side of the equilibrium.'''
 In stable non-dissipative, conservative systems, all perturbations lead to undamped oscillations. Yet in dissipative systems oscillations can get amplified and, for the convective overstability, the relevant criterion is that the Prandtl number is significantly smaller than 1 but sufficiently larger than 0.  In other words, one needs thermal conductivity or equivalently heat transport by radiation that occurs on timescales of the dynamical system, while at the same time acting much more efficient than the viscosity of the underlying fluid.

{The destabilizing influence of a finite thermal time on epicyclic oscillations we are describing in this paper is analogous to the usual heat engine explanation of the $\kappa$ and $\epsilon$ mechanisms in stars \citep{Eddington_1926,Cox_1980}. In the anelastic approximation, if the radial pressure gradient is negative, a fluid element undergoing epicyclic oscillations experiences a negative Lagrangian temperature perturbation ($\delta T$,
measured with respect to the element's initial temperature) when it is displaced outward, and a positive one when displaced inward: i.e.~$\delta T \delta R < 0$. If the radial gradient of entropy is also negative, however, then the
Eulerian temperature perturbation ($T_1$, measured with respect to the background gas) has the opposite sign to $\delta T$, so that the element loses heat to its surroundings when $\delta T < 0$ and gains it when $\delta T > 0$. Thus, if the entropy of the element returns to its original value after a complete oscillation, the element rejects less heat during the outward half of the cycle than it absorbs during the inward part. The difference appears as an increase in the mechanical energy of the oscillation.}

As a linear overstability, radial convection offers a route to angular momentum transport and vortex formation
even in disks without adequate ionization to support the MRI.
The vortices require a negative radial entropy gradient, and their growth rates are quite small.
{Yet based on the two dimensional {radial-vertical} runs presented in this paper it is not obvious how the sheetlike (vertically thin) modes could develop nonlinearly into vortices that are coherent structures across a vertical pressure scale height, nor can we right now determine whether these modes should be effective in transporting angular momentum outward: the instability feeds off the entropy gradient rather than
orbital energy, so the direction of angular momentum transport is not constrained by energy considerations.
Both questions will have to be addressed in 3D simulations currently in preparation.}

At the same time other hydrodynamical instabilities may occur in dead zones of accretion disks.
One class of instability has its sweet spot for very short cooling times or vertically adiabatically stratified disk,
e.g.\ the Goldreich-Schubert-Fricke (G.S.F.) instability named after the work by \citet{Goldreich67} and \citet{Fricke68}. For a recent discussion of the role of the GSF in accretion disks see the work by \citet{Nelson13}. Another instability prefers vertically isothermal disks with as little cooling as possible, e.g.\ ``critical-layer-instability'' \citep[C.L.I.,][]{Marcus13}. Thus the G.S.F. and the C.L.I are mutually exclusive, whereas
the convective observability falls in the middle of the parameter range with respect to the cooling time, and has no preference on the vertical 
stratification of the disk. 
One can now easily extrapolate that, much like the case in complicated climate systems such as the Earth's atmosphere, there is not just one single instability
responsible for all weather phenomena, but rather a zoo of instabilities that operate both independently and hand-in-hand.
Future work on hydrodynamical (as opposed to magneto-hydrodynamical) instabilities and overstabilities in protoplanetary disks
has to focus on two questions: 1.) How do the three above mentioned instabilities interact and how far from their sweet spot can they still operate; and 2.) What is the actual occurring range of vertical and radial stratification plus cooling efficiencies in 
protoplanetary disks.

\acknowledgments
Our special thanks to Peter Bodenheimer for keeping our morale high.
The work of A.H. was in part funded by a fellowship from the Alexander von Humboldt foundation.
The remainder of his work was supported by National Science Foundation, Cyberenabled Discovery Initiative grant AST08-35734, 
National Aeronautics and Space Administration grant NNX10AI42G (DSE), 
and a Kalbfleisch Fellowship from the American Museum of Natural History.

\appendix

\section{Numerical Simulations}
We used the Tramp Code as described in \citet{KHK99} and \citet{KB03} to perform
2D axissymmetric cylindrical simulations of a vertically unstratified accretion disk.
In order to allow for the vortical modes associated with the over-stability we need {at least two degrees} of freedom. 
The second dimension could in principle be the azimuthal direction, but {based on simulations in the past
\citet{Petersen07} and our considerations in Section 2.1 our derived growth behavior does not
hold for vertically integrated disk simulations.}
%
%
%
We therefore take advantage of the property of the convective overstability that it works in axisymmetric geometries,
 unlike some other instabilities such as the baroclinic instability.
Vertical gravity was excluded from the simulations {as it is not included in our analytic work as well. Vertical gravity in combination with a radial temperature gradient} would lead to vertical shear and thus make it difficult to separate between effects due to the G.S.F.-instability and the convective overstability.

We used a cylindrical coordinate system that
 spans from $R_{in} = 0.8$ to $R_{out} = 1.2$ and from $z_{-} = -0.2$ to $z_{+} = -0.2$. 
The grid has constant spacing and for the production runs we chose the number of grid cells in the radial and vertical direction to be NX = 256 and NY = 256. The initial density and temperature profile follow radial power laws, set by p and q (see above). The pressure scale height at R = 1 is chosen to be H/R = 0.1, and has a radial powerlaw determined by q. The adiabatic index $\gamma$ for 3D  solar nebula gas is $\gamma = 1.43$. The azimuthal velocity is initialized at the pressure supported sub-Keplerian value 
\begin{equation}
v_\phi = \Omega(R) R \sqrt{1 + \left(H/R\right)^2 \left(p + q\right)}.
\end{equation}
The initial velocities in radial, vertical and azimuthal directions are randomly perturbed at a level of $\pm 10^{-8}$ or
$\pm 10^{-4} c_s$.
We have reflecting boundary conditions in the radial direction and periodic boundary conditions
in the vertical direction. To minimize the impact of the radial boundary conditions, we only use data from the inner
$50\%$ of the simulation volume: $0.9 \le R \le 1.1$.
\begin{figure}
\includegraphics[width=0.5\columnwidth]{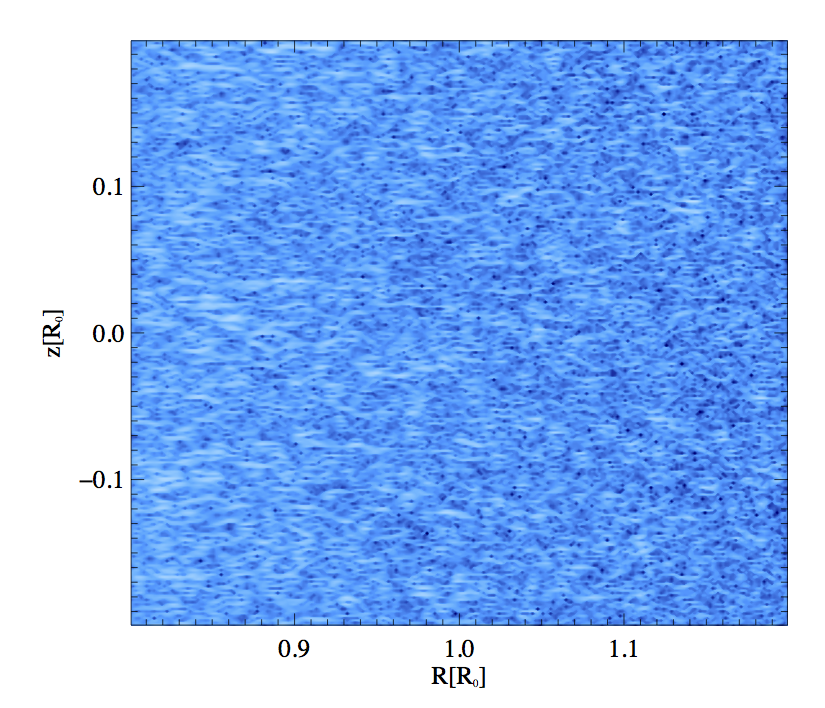}
\includegraphics[width=0.5\columnwidth]{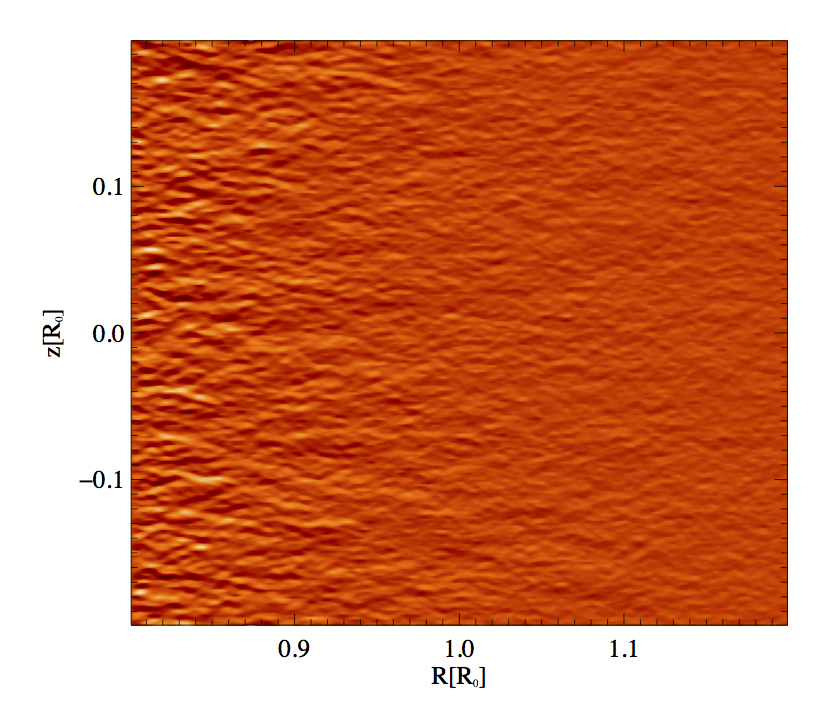}
\includegraphics[width=0.5\columnwidth]{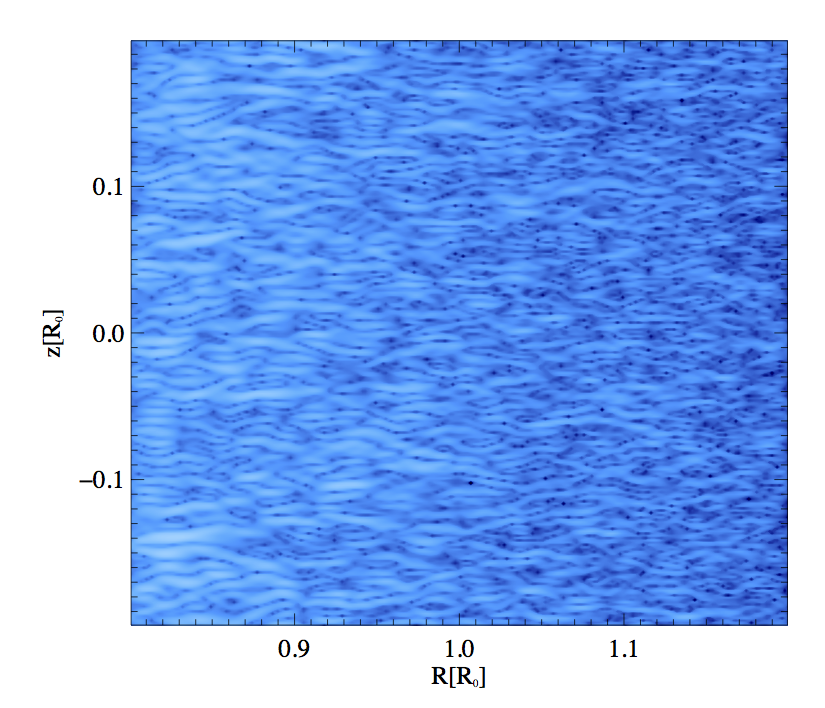}
\includegraphics[width=0.5\columnwidth]{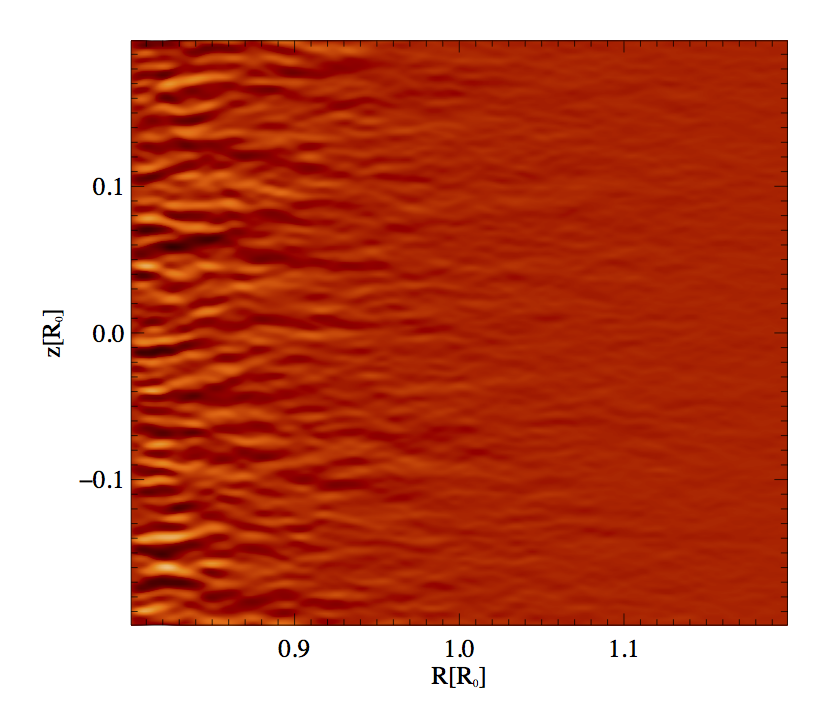}
\caption{Evolution of the instability for $q=-2;p=-1.32$: From top to bottom at 13 and 65 orbits after initial perturbation.
On the left (blue) we plot the absolute value of the velocity in units of the local sound speed and on the right the relative perturbations of the local temperature (red) against the back ground temperature. The colors are scaled to the actual maximum value for better contrast: $v_{\rm max}(t=13, 65)/c_s = 0.0013, 0.015; dT_{\rm max}(t=13, 65)/T_0 = 0.00012, 0.0011$.
\label{Fig_thov4.png}}
\end{figure}
\begin{figure}
\includegraphics[width=0.75\columnwidth]{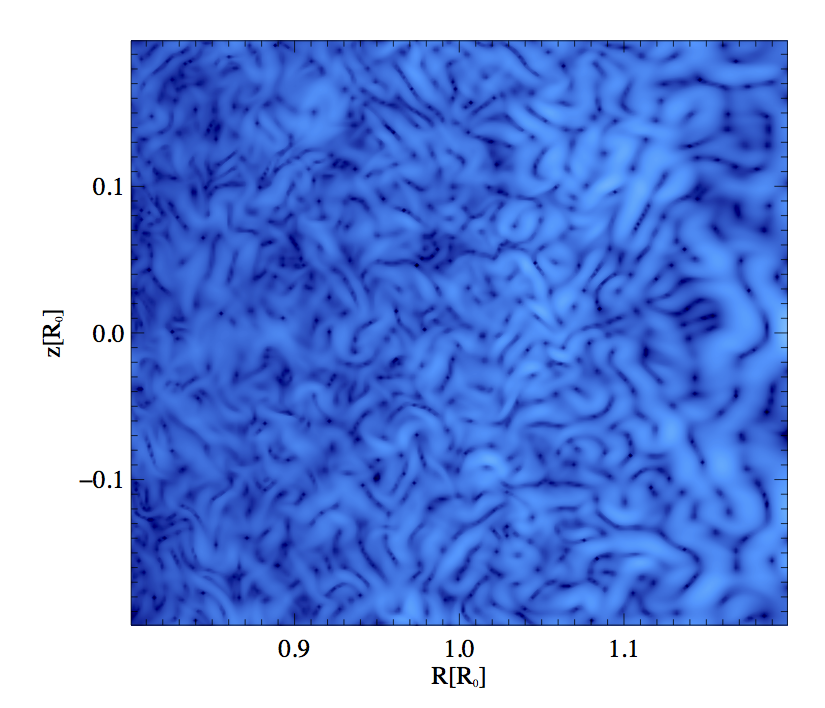}
\includegraphics[width=0.75\columnwidth]{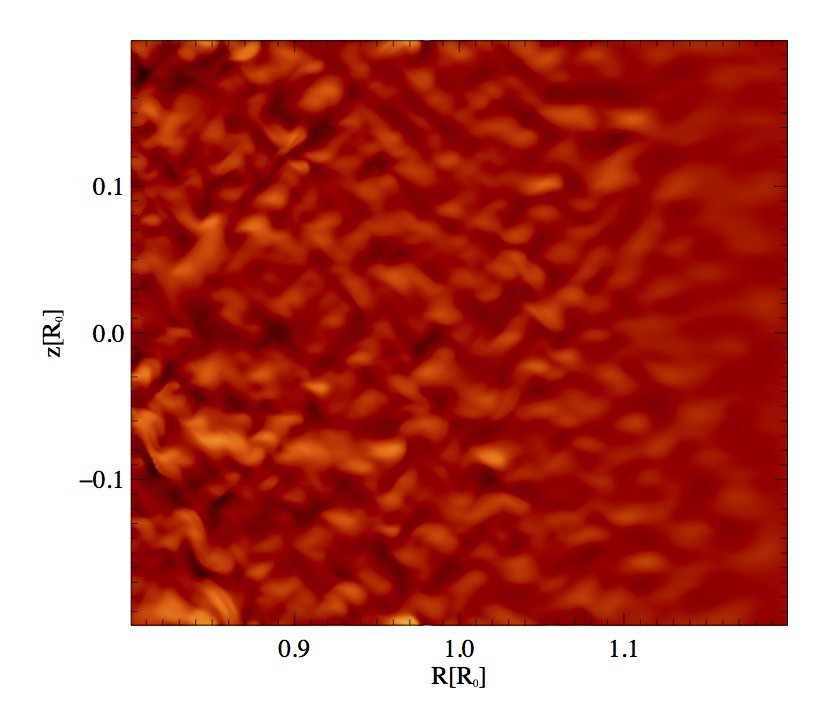}
\caption{Evolution of the instability for $q=-2;p=-1.32$: 130 orbits after initial perturbation.
Colors same as Fig.  \ref{Fig_thov4.png}: $v_{\rm max}(t=130)/c_s = 0.045; dT_{\rm max}(t=130)/T_0 =  0.0036$.
\label{Fig_thov5.png}}
\end{figure}
In Figs.\ \ref{Fig_thov4.png} and \ref{Fig_thov5.png} we show three snapshots of the evolution of the instability for the most unstable case $q=-2; p=-1.32; \gamma\Omega\tau = 1$. One taken at 13 orbits after the perturbation one after 65 and one after 130 orbits (from top to bottom).
On the left we plot the absolute value of the velocity in units of the local sound speed and on the right the relative perturbations of the local temperature against the mean back ground temperature.
{The density fluctuations have generally the same amplitude as the temperature fluctuations, whereas the measured pressure fluctuations are an order of magnitude smaller than the adiabatic pressure fluctuations corresponding to the density fluctuations. This observation is an a posteriori verification of our quasi-hydrostatic approach.}
The initial perturbations grow both in amplitude as well as in spatial extent.
Test simulations without thermal relaxation or without unstable radial stratification did not show any growth.


\begin{thebibliography}{}

\bibitem[Andrews et al.(2010)]{andrews10} Andrews, S.~M., Wilner, 
D.~J., Hughes, A.~M., Qi, C., Dullemond, C.~P.\ 2010.\ Protoplanetary Disk 
Structures in Ophiuchus. II. Extension to Fainter Sources.\ The 
Astrophysical Journal 723, 1241 

\bibitem[Bai(2011)]{2011ApJ...739...50B} Bai, X.-N.\ 2011, \apj, 739, 50 


\bibitem[Balbus \& Hawley(1991)]{BH91} Balbus, S.~A., \& Hawley, J.~F.\ 1991, \apj, 376, 214

\bibitem[Bell et al.(1995)]{bell95} Bell, K.~R., Lin, 
D.~N.~C., Hartmann, L.~W., \& Kenyon, S.~J.\ 1995, \apj, 444, 376 

\bibitem[Bell et al.(1997)]{bell97} Bell, K.~R., Cassen, 
P.~M., Klahr, H.~H., \& Henning, T.\ 1997, \apj, 486, 372

\bibitem[Cox(1980)]{Cox_1980} Cox, J.~P.\ 1980, Research 
supported by the National Science Foundation Princeton, NJ, Princeton 
University Press, 1980.~393 p.,  


\bibitem[Eddington(1926)]{Eddington_1926} Eddington, A.~S.\ 1926, The 
Internal Constitution of the Stars, Cambridge: Cambridge University Press, 
1926.~ ISBN 9780521337083.,  





\bibitem[D'Alessio et al.(2005)]{D'Alessio05} D'Alessio, P., 
Calvet, N., 
\& Woolum, D.~S.\ 2005, Chondrites and the Protoplanetary Disk, 341, 353 

\bibitem[Fricke(1968)]{Fricke68} Fricke, K.\ 1968, \zap, 68, 317 


\bibitem[Gammie(1996)]{Gammie96} Gammie, C.~F.\ 1996, \apj, 457, 
355 

\bibitem[Goldreich 
\& Schubert(1967)]{Goldreich67} Goldreich, P., \& Schubert, G.\ 1967, \apj, 150, 571 


\bibitem[Kim \& Ostriker(2000)]{2000ApJ...540..372K} Kim, W.-T., \& Ostriker, E.~C.\ 2000, \apj, 540, 372 


\bibitem[Klahr et al.(1999)]{KHK99} Klahr, H.~H., Henning, 
T., \& Kley, W.\ 1999, \apj, 514, 325 

\bibitem[Klahr 
\& Bodenheimer(2000)]{KB00} Klahr, H., \& Bodenheimer, P.\ 2000, Disks, Planetesimals, and Planets, 219, 63 

\bibitem[Klahr 
\& Bodenheimer(2003)]{KB03} Klahr, H.~H., \& Bodenheimer, P.\ 2003, \apj, 582, 869 

\bibitem[Klahr(2004)]{K04} Klahr, H.\ 2004, \apj, 606, 1070 

\bibitem[Klahr et al.(2013)]{KRL13} Klahr, H., Raettig, N., 
\& Lyra, W.\ 2013, European Physical Journal Web of Conferences, 46, 4001 

\bibitem[Kley et 
al.(2009)]{bitsch09} Kley, W., Bitsch, B., \& Klahr, H.\ 2009, \aap, 506, 971 


\bibitem[Lesur 
\& Papaloizou(2010)]{LP2010} Lesur, G., \& Papaloizou, J.~C.~B.\ 2010, \aap, 513, A60 

\bibitem[Lyra 
\& Klahr(2011)]{LK2011} Lyra, W., \& Klahr, H.\ 2011, \aap, 527, A138 

\bibitem[Marcus et al.(2013)]{Marcus13} Marcus, P.~S., Pei, S., 
Jiang, C.-H., 
\& Hassanzadeh, P.\ 2013, Physical Review Letters, 111, 084501 

\bibitem[Miller 
\& Stone(2000)]{2000ApJ...534..398M} Miller, K.~A., \& Stone, J.~M.\ 2000, \apj, 534, 398 


\bibitem[Nelson et al.(2013)]{Nelson13} Nelson, R.~P., Gressel, 
O., \& Umurhan, O.~M.\ 2013, \mnras, 435, 2610 

\bibitem[Perez-Becker 
\& Chiang(2011)]{2011ApJ...727....2P} Perez-Becker, D., \& Chiang, E.\ 2011, \apj, 727, 2 


\bibitem[Petersen et al.(2007)]{Petersen07} Petersen, M.~R., 
Julien, K., \& Stewart, G.~R.\ 2007, \apj, 658, 1236 

\bibitem[Raettig et al.(2013)]{RLK13} Raettig, N., Lyra, W., 
\& Klahr, H.\ 2013, \apj, 765, 115 

\bibitem[Shakura 
\& Sunyaev(1973)]{SS73} Shakura, N.~I., \& Sunyaev, R.~A.\ 1973, \aap, 24, 337 


\bibitem[Turner et al.(2010)]{2010ApJ...708..188T} Turner, N.~J., 
Carballido, A., \& Sano, T.\ 2010, \apj, 708, 188 

\bibitem[R{\"u}diger et al.(2002)]{Rudiger02} R{\"u}diger, G., Arlt, R., \& Shalybkov, D.\ 2002, \aap, 391, 781 

\bibitem[Velikhov(1959)]{Vel59} Velikhov, E. P.,\ 1959, Sov. Phys. JETP, vol. 9 (1959), pp. 995�998 (in Russ.).

\end{thebibliography}
\end{document}